\documentclass{ws-acs}
\usepackage{url}
\usepackage{xcolor}

\begin{document}

%\makeatletter
%\def\@biblabel#1{[#1]}
%\makeatother

%\catchline{}{}{}{}{}

\title{On the origins of extreme wealth inequality\\in the Talent vs Luck Model}

\author{DAMIEN CHALLET }

\address{Universit\'e Paris Saclay, CentraleSup\'elec, Laboratoire de Math\'ematiques et Informatique pour la Complexit\'e et les Syst\`emes, 9 Rue Joliot-Curie, 91190 Gif-Sur-Yvette, France\\
damien.challet@centralesupelec.fr}

\author{ALESSANDRO PLUCHINO }

\address{Department of Physics and Astronomy, University of Catania, and INFN - CT, \\Via S.Sofia 64, Catania 95123, Italy\\
alessandro.pluchino@ct.infn.it}

\author{ALESSIO EMANUELE BIONDO}

\address{Department of Economics and Business, University of Catania, \\Corso Italia 55, Catania 95129, Italy\\
ae.biondo@unict.it}

\author{ANDREA RAPISARDA}

\address{Department of Physics and Astronomy, University of Catania, and INFN - CT, \\Via S.Sofia 64, Catania 95123, Italy\\ Complexity Science Hub Vienna, Austria\\
andrea.rapisarda@ct.infn.it}

\maketitle

%\begin{history}
%\received{(received date)}
%\revised{(revised date)}
%\accepted{(Day Month Year)}
%\comby{(xxxxxxxxxx)}
%\end{history}

\begin{abstract}
While wealth distribution in the world is highly skewed and heavy-tailed, human talent -- as the majority of individual features -- is normally distributed. In a recent computational study by Pluchino {\it et al} [Talent vs luck: The role of randomness in success and failure, {\it Adv. Complex Syst.} {\bf 21} (03-04) (2018) 1850014], it has been shown that the combined effects of both random external factors (lucky and unlucky events) and multiplicative dynamics in capital accumulation are able to clarify this apparent contradiction. We introduce here a simplified version (STvL) of the original Talent versus Luck (TvL) model, where only lucky events are present, and verify that its dynamical rules lead to the same very large wealth inequality as the original model. We also derive some analytical approximations aimed to capture the mechanism responsible for the creation of such wealth inequality from a Gaussian-distributed talent. Under these approximations, our analysis is able to reproduce quite well the results of the numerical simulations of the simplified model in special cases. On the other hand, it also shows that the complexity of the model lies in the fact that lucky events are transformed into an increase of capital with heterogeneous rates, which yields a non-trivial generalization of the role of multiplicative processes in generating wealth inequality, whose fully generic case is still not amenable to analytical computations. 
\end{abstract}


\section{Introduction}

It is well known that the wealth distribution in the world is highly skewed, with wealth being concentrated in the hands of a very small number of very rich individuals. Recently, this gap has further increased: just eight men own the same wealth as the poorest half of humanity, i.e. about 3.6 billion people \cite{Hardoon}. More generally, as originally discovered by Pareto \cite{Pareto}, the tail of wealth distribution follows a heavily-tailed power law distribution where 80\% of people of a country own only the 20\% of the total capital and the 20\% richest own the remaining 80\% (the precise figures vary in time). These intriguing features of the wealth distribution have been largely studied in the last decades through many theoretical models developed in the context of statistical physics, game theory and complex networks theory. Two broad categories emerge: models where multiplicative processes and redistribution are responsible for wealth distributions with a power-law tail \cite{Bouchad,Garlaschelli,Fiaschi} and exchange models producing a Gamma distribution \cite{Angle,Patriarca} (see \cite{LuxAngle} for a critical review). 

On the other hand, it is equally well known that the human talent is normally distributed among a population \cite{Wechsler,Kaufman1,Kaufman2,Stewart}; the same applies for the efforts which an individual can invest during each single week of her life in the attempt of achieving success \cite{Erickson}. Finally, it is also accepted that randomness (good or bad luck) plays a not negligible role in determining the outcome of our efforts: for example, living in an environment rich of opportunities or being in the right place at the right time, are considered to be decisive incentives for becoming rich or successful \cite{Milanovic}. But, again, fortune is blind by definition, thus, assuming the same external conditions, one should not expect, in principle, extreme differences in the occurrence of either lucky or unlucky events among the individuals in the everyday life. All such considerations done, why is the wealth (i.e., the success) so unevenly distributed, given that talent and luck are much less differentiated than it? 

In order to answer this question, three of us recently introduced an agent-based model called "Talent vs Luck" \cite{Pluchino1} (TvL thereafter), which leads to a heavy-tailed distribution of capital in a population of individuals, despite of the non heavy- tailed distributions of both talent and random events (positive and negative). In the original TvL model, individuals are endowed with a normally distributed talent and the same initial amount of capital, and are exposed to the random action of both positive/lucky and negative/unlucky events. When a lucky event occurs, a person doubles the capital with a probability equal to her talent; contrariwise, an unlucky event halves her capital with certainty. 

At the end of the simulation, as a result of such a multiplicative dynamics, the probability distribution of capital is heavy-tailed: approximately a power-law with a negative exponent between 1 and 2 \cite{Pluchino1}. This seems to mimic the well-known ''Mathew effect'' or ''rich get richer'' effect, induced by the feedback mechanisms of the real socio-economic complex networks \cite{Merton}. The point of that model is that success and talent appeared to be not much correlated, as shown for example by the fact that richest individuals almost never were the most talented ones. In other words, very lucky people, although moderately talented, appeared to have much more possibilities to reach the apex of the social success than very talented but unlucky persons ---a finding in agreement with our perception of real life.

In this paper we explore the origin of the extreme wealth/success inequality in the TvL model by considering a simplified version of the latter, that hereafter we call STvL model, where, in particular, the interplay between the distributions of talent and of the number of lucky events is addressed (unlucky events are not present at all).  If the lucky event number distribution in the population was exponential, the multiplicative nature of capital increase would immediately lead to a pure power law-distributed capital and the model would be trivial. We show here that this is not the case, and that the STvL model transforms a non-exponential distribution of number of events into a complex heavy-tailed distribution of capital because of the heterogeneity of talent. Quite notably, the final distribution, while heavy-tailed, has no power law tails in the limits considered. 

We first describe in detail, in Sec.2, the simplified model and show with numerical results that the STvL model is still able to reproduce the main stylized facts of the original model. We also show that a finite number of time steps increases the heaviness of wealth distributions. We then discuss, in Sec.3, the basis of our formal approach and present the analytical derivation of the capital/success inequality, under different assumptions related to the talent distribution among individuals. Finally, in Sec.4, we present some conclusive remarks.

Appendix A is devoted to the analytical extension of the STvL model to the original TvL one, where both lucky and unlucky events are present.

\section{The Simplified TvL Model (STvL)}

\subsection{Model description}

%%%%%%%%%%%%%%%%%%%%%%%%%%
%%%%%%%%%  FIG. 1
\begin{figure}
\begin{center}
\includegraphics[width=2.7in,angle=0]{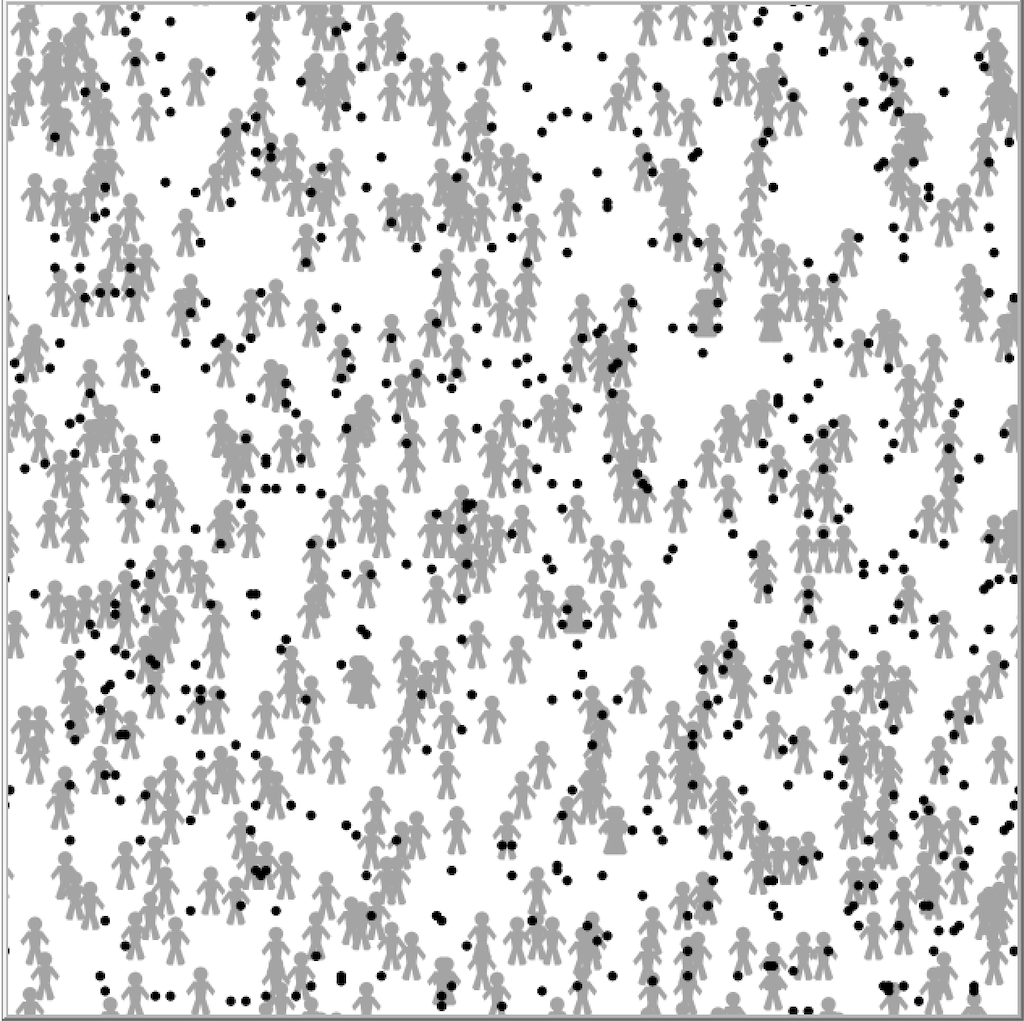}
\caption{\small 
An example of initial setup for a single simulation run. $N=500$ individuals (agents, in gray), with different degrees of talent (intelligence, skills, etc.), are randomly located at their fixed positions within a continuous square world with periodic boundary conditions. During each simulation, they are exposed to $N_E=500$ lucky events (black points) randomly moving across the world \cite{Netlogo}.          
}
\label{world} 
\end{center}
\end{figure}
%%%%%%%%%%%%%%%%%%%%%%%%%%

Let us consider $N$ individuals randomly placed at fixed positions within a square continuous world with periodic boundary conditions and surrounded by a given number $N_E$ of lucky event-points, corresponding to opportunities occurring by chance in the real world. These event-points are also initially randomly placed. Thus, for relatively small values of the ratio  $N_E / N$, at the beginning of each simulation there will be a greater random concentration of event-points in different areas of the world, while other areas will be more neutral. In addition, for a relatively small number of iterations, the random placement of individuals also induces small spatial correlations in the number of lucky events. Individuals and luck event points are modeled as disks of diameter 1 in a 300x300 square (agents and events may partially overlap). Note that a discrete version of the model can be also defined, but it does not change the results qualitatively.

In Fig.\ \ref{world}, an example of world with $N=500$ agents and $N_E=500$ lucky event points is shown. At each time step, lucky events move according to an unbiased random walk (with steps of size 2), which neither depends on the presence of the individuals, nor on their intrinsic qualities. The further random movement of the points inside the world does not change this fundamental feature of the model, which exposes different individuals to different amount of opportunities during their life. Of course, keeping the events fixed and letting the agents move would have been an equally valid solution, but our choice better highlights the active role of luck in choosing who to offer opportunities. 

At the beginning of each simulation run, each agent is endowed with a given level of talent and with an initial capital. The talent of agent $i$ is represented by a real variable drawn in the interval $[0,1]$ from a known symmetric distribution $P(T)$, e.g. a Gaussian $T_i \sim \mathcal{N}(\mu_T,(\sigma_T)^2)$, constant for the whole duration of the simulation, truncated to account for the finite support. Note that we take take a small $\sigma_T$, which makes the truncation practically inexistent. As in the original TvL model, talent is meant to represent any kind of ability (including intelligence, skill, efforts, etc...) which allows an individual to transform a random opportunity into reality. Therefore, having a great/small talent represents a strong a-priori advantage/disadvantage for a given agent. On the other hand, the initial capital $C_i$ of the agents, which represents their starting level of success/wealth (expressed in dimensionless units), is distributed according to a thin-tailed distribution in order to not offer any comparatively large initial advantage to anyone. We use here a uniform distribution of $C\in[0.5,1.5]$ for $P(C)$.
%%%%%%%%%  FIG. 2
\begin{figure}
\begin{center}
\includegraphics[width=3.6in,angle=0]{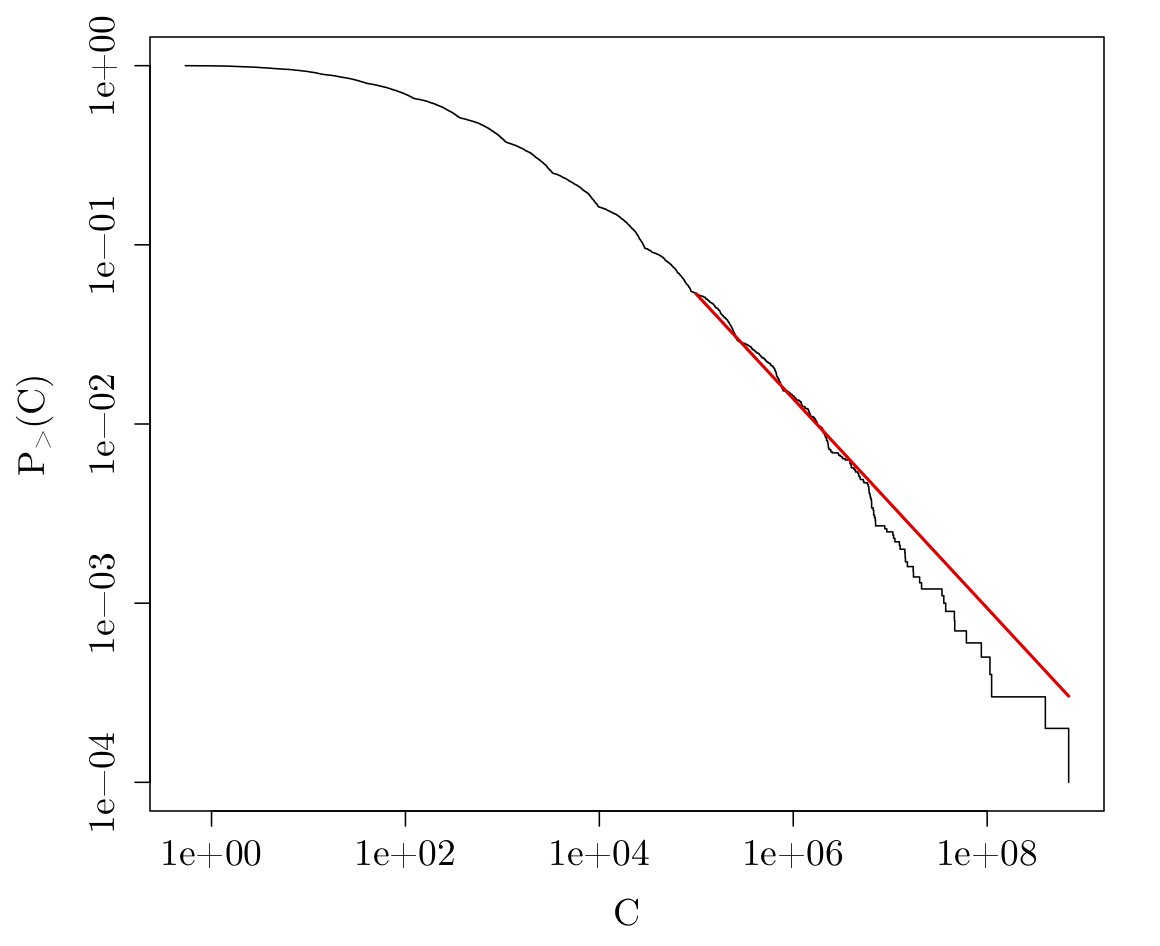}
\caption{\small Reciprocal cumulative probability distribution $P_>(C)$ of capital among the population of $N=10000$ agents with $N_{E}=5000$ lucky event points in log-log scale, obtained after a single simulation run with $M=30$ time steps. Despite the normal distribution of talent and the equally distributed initial capital, the final distribution of capital/success is heavy tailed. Fitting its tail with a power-law \cite{Clauset} yields an exponent $\simeq -1.6$ (red line). 
} 
\label{capital} 
\end{center}
\end{figure}
%%%%%%%%%%%%%%%%%%%%%%%%%%

A single simulation run lasts $M$ time steps, each of which corresponding to the typical duration for lucky players to double their capital, roughly a year, as discussed in \cite{Pluchino1}. In fact, during the time evolution of the model, all event-points randomly move within the world, then, at a given time step, some of them possibly intercept the position of a given agent $A_i$, i.e. lie within a circular neighborhood of radius 1 around that agent. In this case, we say that a lucky event (an opportunity) has occurred at that time step and, as a consequence, agent $A_i$ doubles her capital/success with a probability proportional to her talent $T_i \in[0,1]$, i.e., $C_i(t)=2C_i(t-1) \, \Leftrightarrow \, rand[0,1] < T_i$ (meaning that the agent is smart enough to take advantage of the opportunity).         

We denote with $n_i$ the total number of opportunities experimented by an agent $A_i$ and with $k_i$ the number of those ones successfully transformed into an increase of capital. At the end of each simulation, both these variables result to be distributed among the agents according to the functions $P(n)$ and $P(k)$ respectively. We are interested in studying the final distribution of capital $P(C)$ and its relationship with $P(n)$ and $P(k)$.

%%%%%%%%%  FIG. 3
\begin{figure}
\begin{center}
\includegraphics[width=3.6in,angle=0]{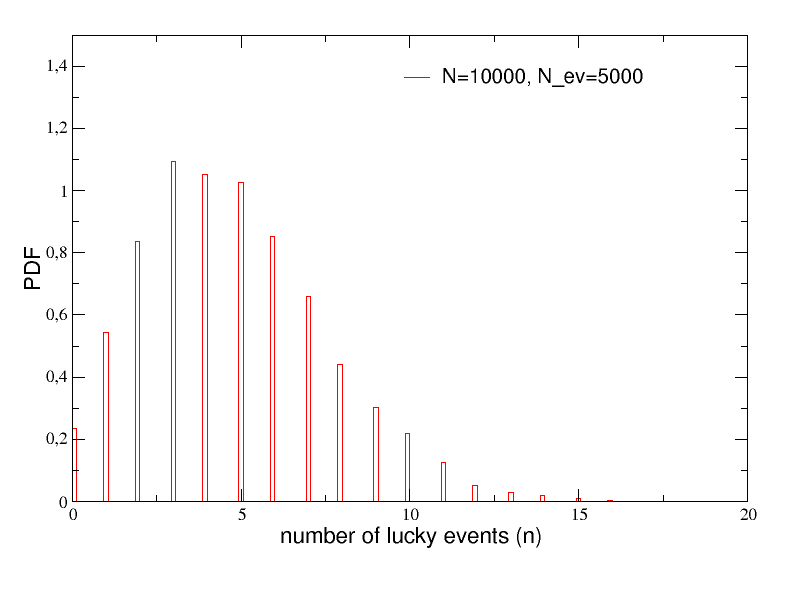}
\caption{Probability distribution $P(n)$ of the number of lucky events occurred to the $N=10000$ agents (and $N_{E}=5000$ lucky event points) during a single run of $M=30$ time steps.}
\label{events-pdf} 
\end{center}
\end{figure}
%%%%%%%%%%%%%%%%%%%%%%%%%%

\subsection{Numerical results}

Consider $N=10000$ agents, with an initial amount of capital $C_i(0)\in[0.5,1.5]$ $\forall i$ and with a talent $T_i \in[0,1]$ following a normal distribution with mean $\mu_T=0.6$ and standard deviation $\sigma_T=0.1$. Further, consider $N_E=5000$ lucky event-points and a time period of $M=30$ simulated time steps. 

At the end of the simulation, as shown in Fig.\ \ref{capital}, we find that the simplified dynamic rules of the STvL model are still able to produce a heavy tailed distribution $P(C)$ of capital/success, with a large amount of  poor (unsuccessful) agents and a small number of very rich (successful) ones. The nature of $P(C)$ is clearly heavy-tailed: to be more precise, assuming that the tail of $P(C)\propto C^{-\alpha}$, i.e., assuming that there is sufficient statistics to produce a clean power-law for small enough $C$ \cite{Voitalov}, we applied the method of \cite{Clauset} and its implementations in R \cite{powerlawR} and Python \cite{powerlawPython} to $P(C)$, which yields $\alpha\simeq1.6$. However, as made clear by Fig.\ \ref{capital}, the tail of $P(C)$ is not a pure power-law, which is confirmed by our analytical approach below. Vuong likelihood ratio tests of a power-law vs a truncated power-law favours the latter (p-value of about 0.0006), while a log-normal distribution is probably better on average than a power-law (p-value of about 0.02). 

As expected, in this simplified model, success and talent are not strongly correlated, success being mostly due to luck. In Fig.\ \ref{events-pdf} the distribution $P(n)$ of the total number $n$ of lucky events occurred to the agents during the same simulation run is reported. It appears quite asymmetric (an effect of the small number of iterations), with a large majority of individuals who experienced a number of events included between $2$ and $10$, while only a very small number of them were so lucky to intercept more than $10$ events. In any case, nobody experienced more than $n_{max}=18$ events. 

%%%%%%%%%  FIG. 4
\begin{figure}
\begin{center}
\includegraphics[width=5.0in,angle=0]{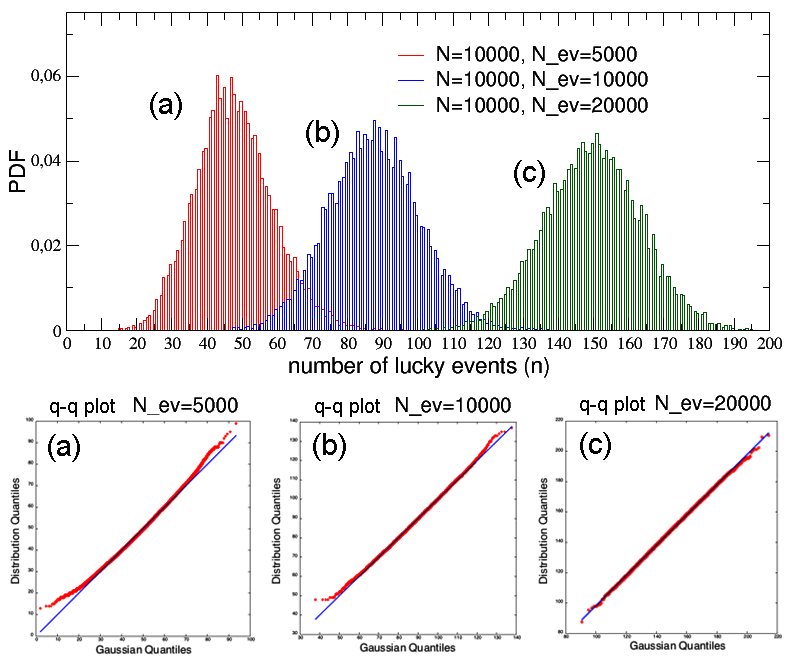}
\caption{\small Top panel: probability distributions $P(n)$ of the number of lucky events occurred to $N=10000$ agents during single runs of $M=300$ time steps and with, respectively, (a) $N_{E}=5000$, (b) $N_{E}=10000$ and (c) $N_{E}=20000$ lucky event points. Bottom panels: the q-q plots of the three distributions shows a convergence to Gaussian behaviour as $N_{E}$ grows, becoming very good for $N_{E}=20000$.
}
\label{luck-pdf} 
\end{center}
\end{figure}
%%%%%%%%%%%%%%%%%%%%%%%%%%

In order to have a clearer idea of how the shape of $P(n)$ depends on the number of lucky event points, we run three simulations with $N=10000$ agents, $N_{E}=5000$, $10000$, and $20000$; in addition, we take $M=300$ time steps in order to be in the steady state. The resulting distributions $P(n)$ are shown in the top panel of Fig.\ \ref{luck-pdf}, while the corresponding q-q plot is reported just below, in the bottom panel of the same figure. It clearly appears that the three distributions progressively tend to assume a Gaussian shape, which becomes very good for $N_{E}=20000$: in this case, the probability of experimenting a lucky event becomes similar for all the agents in the limit of large number of event points. 

It is interesting to notice that, when $M$ is relatively small, the scarcity of opportunities coupled with the fixed spatial positions of the agents reinforces the heaviness of wealth distribution. For very large $M$ those effects disappear and, when the ratio $N_E / N$ is also large enough, $P(n)$ becomes Gaussian. However, even in this case, the heterogeneity of talent guarantees that the resulting capital distribution is non-trivial. We also note that, regrettably, the case $M\to\infty$ is irrelevant in real life in which spatial distribution does matter. In short, while the diffusion of lucky event point is not an essential ingredient for producing non-trivial heavy-tailed capital distributions, it adds relevant spatial correlations in the relevant finite time case of human life.

\begin{figure}
\begin{center}
\includegraphics[width=2.8in,angle=0]{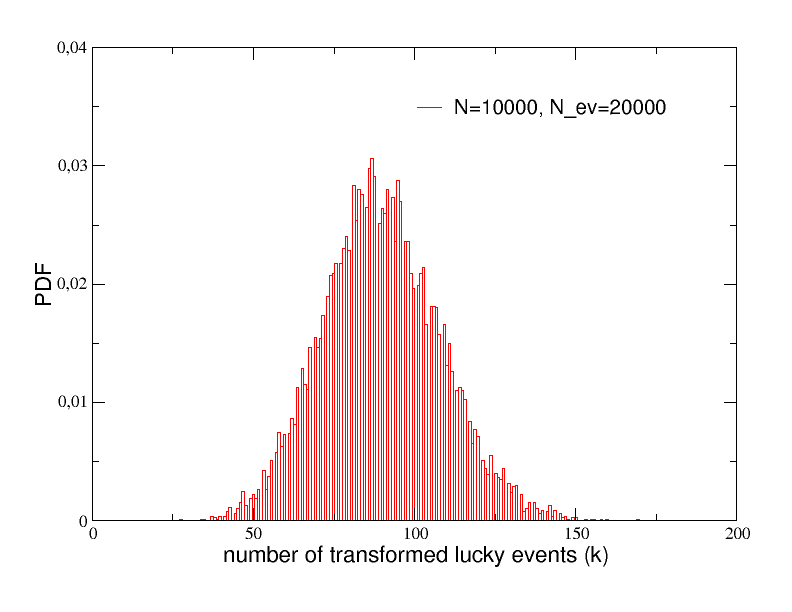}
\includegraphics[width=2.1in,angle=0]{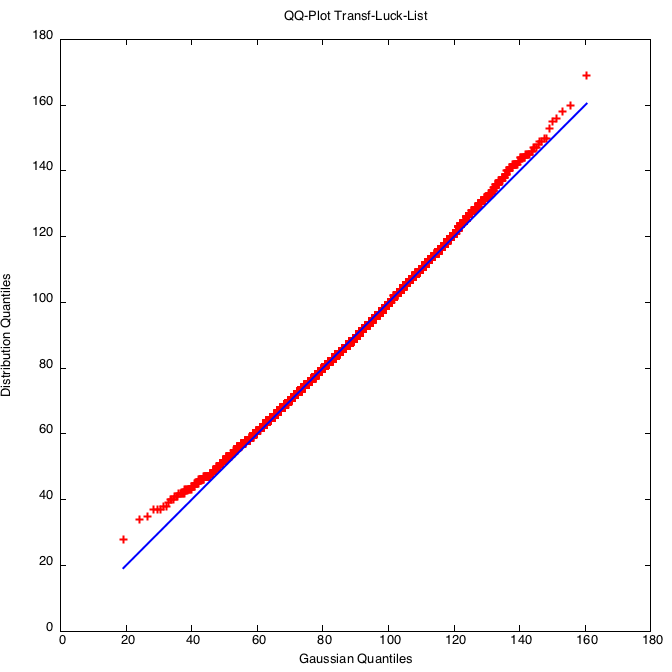}
\caption{Left panel: probability distribution $P(k)$ of the number of transformed lucky events for the $N=10000$ agents during a single run of $M=300$ time steps and with $N_{E}=20000$ lucky event points. Right panel: the q-q plot of the same distribution shows a consistent deviation from normal behavior.
}
\label{transformed-pdf} 
\end{center}
\end{figure}

Finally, Fig.\ \ref{transformed-pdf} reports the distribution $P(k)$ of the transformed opportunities, with $N_{E}=20000$, along with the corresponding q-q plot; a sensible deviation from Gaussian behavior is still observed, as the heterogeneity of $T$ makes $P(k)$ heavier than $P(n)$ in this case. These findings are discussed in details in the next section, where an analytical derivation of $P(C)$ as function of $P(T)$, $P(n)$ and $P(k)$ is presented.  

%%%%%%%%%  FIG. 5
%%%%%%%%%%%%%%%%%%%%%%%%%%

\section{Analytical approach to the STvL model}

The main result of \cite{Pluchino1} was that $P(C)$ has a heavy tail despite the fact that neither $P(T)$ nor $P(n)$ have one. In the previous section we verified that such a feature still holds true also in the simplified version of the model introduced in this work. This section is devoted to the analytical characterization of the relationship between these distributions in the STvL.

\subsection{Formal link between lucky events and capital distributions}

Let us first state a few simple relationships between variables. As we have already seen, during a simulation, agent $i$ experiences $n_i$ lucky events, with each of them being transformed into a capital increase with probability $T_i$, which results into $k_i$ transformed events. Let us drop the indices $i$: $P(k|n,T)$ is therefore a binomial distribution $\mathcal{B}(n,T)(k)$. 

Formally, if $M$ denotes the total number of time steps of a simulation.
\begin{equation}\label{eq:PkT}
P(k|T)=\sum_{n=k}^{M}P(k|n,T)P(n)=\sum_{n=k}^{M}\mathcal{B}(n,T)(k)P(n).
\end{equation}
and thus
\begin{equation}
\label{eq:Pk}
P(k)=\int dT P(T)\sum_{n=k}^{M}\mathcal{B}(n,T)(k)P(n).
\end{equation}

Assuming that all the agents have the same probability to $\rho$ to experience a lucky event in a given time step of the simulation, the number of lucky events of a given agent should follow a binomial distribution, which can be approximated by a Gaussian distribution for large $N$: $P(n)\sim\mathcal{N}(\mu_n,\sigma_n^2)$ with $\mu_n=\rho M$ and $\sigma_n=M\rho(1-\rho)$. As we have seen in the previous section, this is also numerically confirmed by the results of the agent-based simulations reported in Fig.\  4, where $N=10000$ and where $P(n)$ became Gaussian as $N_{E}$ becomes large enough to make $\rho$ constant for all agents (the figure evidently shows that, as $N_{E}$ increases, also $\rho$ increases; the estimate is $\rho = \mu_n / M$, being $\mu_n$ the mean in each of the three reported distributions and $M=300$).  

Assuming that $P(k)$ is known, let us discuss how $P(C)$ may gain its heavy tails. Using $C_i(T)=C_i(0)2^{k_i}$, assuming that $C_i(0)=1$ $\forall i$, that $C$ and $k$ are continuous variables, and dropping the index $i$,
$$
P(C)= P[k=\log_2 C] \frac{dk}{dC}\propto\frac{1}{C}P(\log_2 C).
$$
Thus, for $P(C)$ to have a heavy tail, i.e., to decrease more slowly than any exponential, $P(k)$ must have a tail which decreases more slowly than $e^{-e^k+k}$. This is the case, e.g. both for $P(k)\propto \exp(-\lambda k)$, which leads to $P(C)\propto C^{-\alpha }$, and $P(k)\sim \mathcal{N}(\mu_k,\sigma_k^2)$, which leads to a log-normal $P(C)$. Both distributions are difficult to distinguish, as it is well known \cite{Clauset}. Let us examine in the following a few specific cases.

\subsection{The homogeneous case $T=1$ }

Suppose that all agents have the same talent $T=1$. In this case $k=n$, $P(k)=P(n)$ is Gaussian and thus $P(C)$ is a pure log-normal distribution. In other words, the heavy tails of $P(C)$ are due to the combination of the stochastic nature of the number of lucky events and the multiplicative process which drives capital increases. While this case is relatively trivial, in the model $P(n)\ne P(k)$ when $T<1$ for all agents. 

\subsection{The homogeneous case $0<T<1$, with  constant T}

Let us assume now that all agents still have the same talent, but chosen in the interval $(0,1)$, i.e. $T'$: $P(T)=\delta(T-T')$. It is worth to notice that this special case resembles a situation often occurring in the real world. For example, it is realized when a few individuals with very high and very similar talent are drawn from a larger social group through any kind of selective test or competition (sportive, artistic, for a working place, etc...): in all such cases, the talent of everyone being almost identical, the final success is necessarily mostly a matter of luck.

Since $n$ and thus $k$ are proportional to $M$, let us write $k=\gamma M$ and $n=\nu M$ and take the $M\gg1$ case, which leads to, dropping the prime of $T'$,
\begin{align}
P(\gamma)\simeq & \int_{\gamma}^{1}d\nu P(\gamma|\nu,T)P(\nu),
\label{eq:P_kappa}
\end{align}
where $P(\gamma|\nu,T)\simeq\mathcal{N}[\nu T,\nu T(1-T)/M](\gamma)$ and $P(\nu)\simeq\mathcal{N}[\rho,\rho(1-\rho)/M](\nu)$. Using the characteristic functions of these two distributions simplifies much the computations:

\begin{align*}
P(\gamma) & \simeq \int_{\gamma}^{1}d\nu\int_{-\infty}^{\infty}\frac{ds}{2\pi}\frac{dt}{2\pi}e^{-is\gamma}e^{isT\nu-\frac{1}{2}\frac{\nu T(1-T)}{M}s^{2}}\times\\
&~~~~~~~~~~~~~~~~~~~~~~~~~~~~e^{-it\nu}e^{it\rho-\frac{1}{2}\frac{\rho(1-\rho)}{M}t^{2}}\\
 & =\int_{-\infty}^{\infty}\frac{ds}{2\pi}\frac{dt}{2\pi}e^{-is\gamma}e^{it\rho-\frac{1}{2}\frac{\rho(1-\rho)}{M}t^{2}}\times\\
&~~~~~~~~~~~~~~~~ \int_{\gamma}^{1}d\nu e^{isT\nu-\frac{1}{2}\frac{\nu T(1-T)}{M}s^{2}}e^{-it\nu}.
\end{align*}

The integral on $\nu$ can be readily performed as only linear terms
in $\nu$ appear in the exponential:
\begin{align*}
\int_{\gamma}^{1}d\nu e^{\nu\left[isT-it-\frac{1}{2}\frac{T(1-T)}{M}s^{2}\right]} & =\frac{1}{\lambda}\left(e^{\lambda}-e^{\gamma\lambda}\right),
\end{align*}
where $\lambda =isT-\frac{1}{2}\frac{T(1-T)}{M}s^{2}-it$.

Setting
\begin{align*}
f(t,\gamma) & =\frac{-1}{2\pi i}\frac{e^{i(\rho-\gamma)t-\frac{1}{2}\frac{\rho(1-\rho)}{M}t^{2}}}{t-sT-\frac{i}{2}\frac{T(1-T)}{M}s^{2}},
\end{align*}

$P(\gamma)$ can be shortened to

\begin{align}
P(\gamma) & \simeq \int_{-\infty}^{\infty}\frac{ds}{2\pi}e^{-is\gamma}e^{isT-\frac{1}{2}\frac{T(1-T)}{M}s^{2}}\int_{-\infty}^{\infty}dtf(t,1)\nonumber \\
 & -\int_{-\infty}^{\infty}\frac{ds}{2\pi}e^{-is\gamma}e^{is\gamma\left(T-\frac{1}{2}\frac{T(1-T)}{M}s^{2}\right)}\int_{-\infty}^{\infty}dtf(t,\gamma).
 \label{eq:Pk_sum}
\end{align}

By Cauchy's theorem, as $f(t,\gamma)$ has only one pole at $t^{*}(s)=sT+\frac{i}{2}\frac{T(1-T)}{M}s^{2}$, thus

\[
\int_{\Gamma}dtf(t,\gamma)=\left.e^{it(\rho-\gamma)-\frac{1}{2}\frac{\rho(1-\rho)}{M}t^{2}}\right|_{t=t^{*}(s)}
\]

if $\Gamma$ encloses $t^{*}(s)$ in an anticlockwise way. Let us
assume that $\Gamma$ is the union of a line $t\in[-a,a]\in\mathbb{R}$
and of the anti-clockwise semi-circle $ae^{i\pi x}$, $x\in[0,1]$,
denoted by $\text{Arc}(a)$. For each $s$, $\Gamma$ encloses $t^{*}(s)$
if $a>|t^{*}(s)|$. Because of the term $e^{-\frac{\rho(1-\rho)^{2}}{2M}t^{2}}$
in the integral, $\lim_{a\to\infty}|f(t\in\text{Arc}(a))|=0$. Hence,

\[
\int_{\Gamma}dtf(t,\gamma)=\int_{-\infty}^{\infty}dtf(t,\gamma)=\left.e^{it(\rho-\gamma)-\frac{1}{2}\frac{\rho(1-\rho)}{M}t^{2}}\right|_{t=t^{*}(s)}.
\]

which is an exponential of a fourth-degree polynomial. The right-hand-side of Eq. (\ref{eq:Pk_sum}) is therefore a sum of two integrals of exponentials of fourth-degree polynomials. 
Let us write the argument of the exponential of the second line of Eq.\ (4) as $ia_{1}s+a_{2}s^{2}+ia_{3}s^{3}+a_{4}s^{4}$. A straightforward  computation yields 

\begin{align*}
a_{1} & =T(\rho-\gamma)-\gamma+\gamma T=T\rho-\gamma T-\gamma=T\rho-\gamma-\gamma T\\
a_{2} & =-\frac{1}{2}\frac{\rho T(1-\rho T)}{M}\\
 a_{3} & =-\frac{1}{2}\frac{\rho(1-\rho)T^{2}(1-T)}{M^{2}}\\
a_{4} & =\frac{1}{8}\frac{\rho(1-\rho)T^{2}(1-T)^{2}}{M^{3}}
\end{align*}

The polynomial in the exponential of the first line of Eq. (\ref{eq:Pk_sum}) has the same $a_3$ and $a_4$, while its first to two coefficients are

\begin{align*}
a_{1}  =&T(\rho-1)-\gamma+T=T\rho-\gamma\\
a_{2} =&-\frac{1}{2}\frac{\rho T(1-\rho T)}{M}
\end{align*}

%%%%%%%%%  FIG. 6
\begin{figure}
\begin{centering}
\includegraphics[width=2.8in,angle=0]{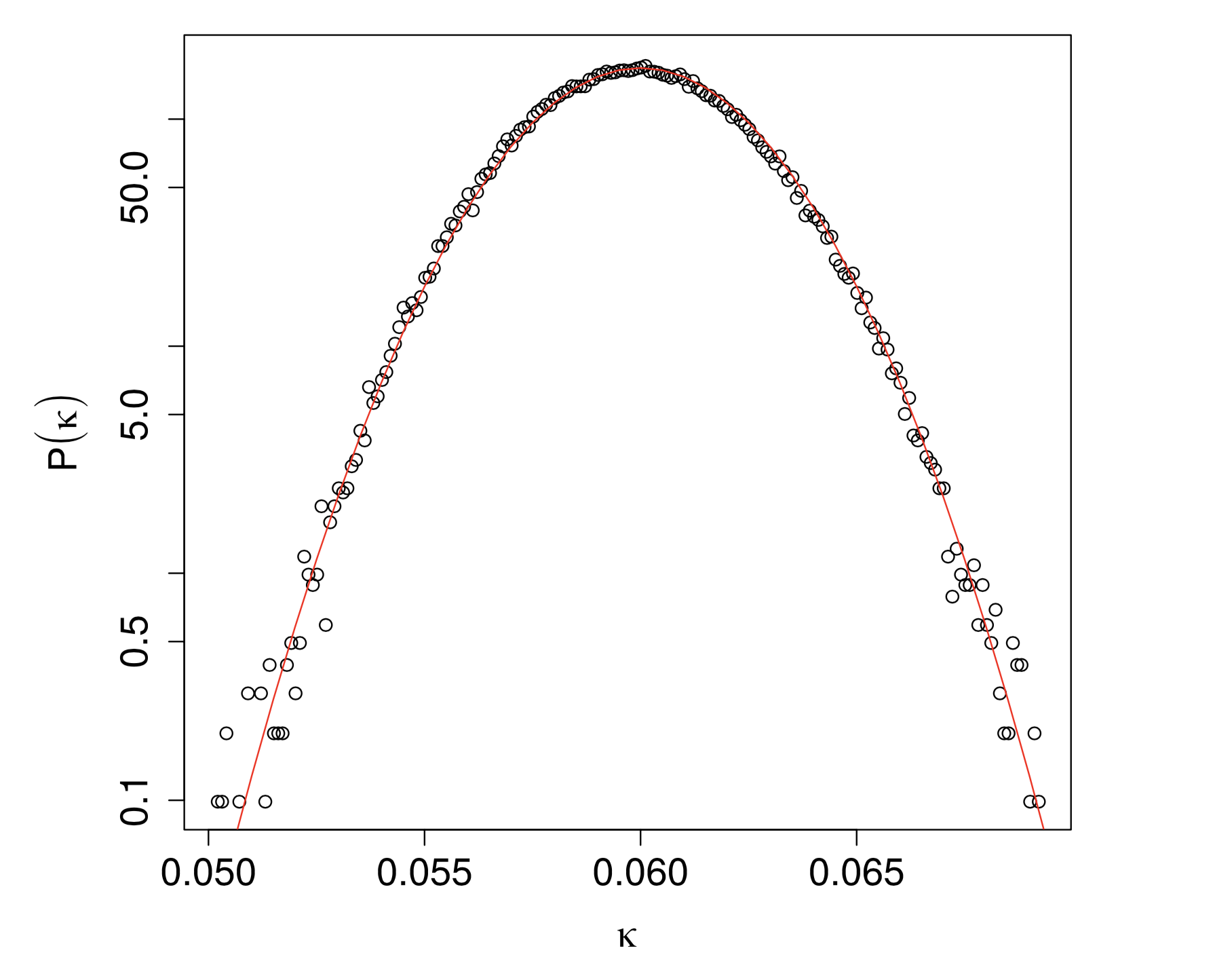} 
\par\end{centering}
\caption{\label{fig:Empirical-vs-theoretical_Pkappa}
Homogeneous case: numerical (circles) vs theoretical (red continuous line) $P(\gamma)$; $T=0.6$, $\rho=0.1$
and $M=10^{5}$; numerical results have been obtained by Monte Carlo simulations of Eq. (\ref{eq:Pk_sum}).}
\end{figure}
%%%%%%%%%%%%%%%%%%

Thus, as expected, the distribution is a Gaussian plus corrections
due to the third and fourth degree terms in the exponential. For large
$M$, these terms have a vanishing influence for small $s$, i.e, for the tails of $P(\gamma)$. Let us therefore neglect them:

\begin{align}
P(\gamma)  \simeq&\int_{-\infty}^{\infty}\frac{ds}{2\pi}e^{-is\gamma}e^{iT\rho s-\frac{1}{2}\frac{\rho T(1-\rho T)}{M}s^{2}}\\&-\int_{-\infty}^{\infty}\frac{ds}{2\pi}e^{-is\gamma}e^{iT(\rho-\gamma)s-\frac{1}{2}\frac{\rho T(1-\rho T)}{M}s^{2}}\nonumber\\
 =&\mathcal{N}\left(\rho T,\frac{\rho T(1-\rho T)}{M}\right)(\gamma)\\&-\frac{1}{1+T}\mathcal{N}\left(\rho\frac{T}{1+T},\frac{\rho T(1-T\rho)}{M(1+T)^{2}}\right)(\gamma)\nonumber.
\end{align}

Fig.\ \ref{fig:Empirical-vs-theoretical_Pkappa} shows that this
is a very good approximation. Thus, once again, the tails $P(k)$ are essentially Gaussian, and $P(C)$ is a log-normal distribution when the probability of lucky event occurence per time step is the same for all the agents. In practice, as shown by Fig.\ \ref{luck-pdf}, when $M$ and $N_E$ are relatively small, this hypothesis does not fully hold, which gives heavier tails to $P(n)$.

\subsection{Heterogeneous case}

The complexity of the model lies in the heterogeneity of talent, which leads to non-trivial distributions.  Let us therefore generalize Eq. (\ref{eq:Pk_sum}) by averaging $T$ over its distribution.  The aim of the original Talent vs Luck model, as well as that of the simplified STvL one, is to show that a thin-tailed distribution of $T$ leads to heavy-tailed $P(C)$ and accordingly uses a Gaussian distribution for $T$, $\mathcal{N}(\mu_{T},\sigma_{T}^{2})$ with a small $\sigma_T$. Analytical computations however, are much simpler for a uniform distribution.
Let us take a simple uniform distribution of $T$ over $[T_0-\frac{a}{2},T_0+\frac{a}{2}]$: 

\begin{align}
P(\gamma)\simeq & \int_{-\infty}^{\infty}\frac{ds}{2\pi}e^{-is\gamma}a\int_{T_{0}-\frac{a}{2}}^{T_{0}+\frac{a}{2}}dTe^{iT\rho s-\frac{1}{2}\frac{\rho T(1-\rho T)}{M}s^{2}}\nonumber\\
 & -\int_{-\infty}^{\infty}\frac{ds}{2\pi}e^{-is\gamma}a\int_{T_{0}-\frac{a}{2}}^{T_{0}+\frac{a}{2}}dTe^{iT(\rho-\gamma)s-\frac{1}{2}\frac{\rho T(1-\rho T)}{M}s^{2}}.\label{eq:P_gamma_unif}
\end{align}

The relevant case is the small heterogeneity limit ($a\ll1$), in which case
%\begin{widetext}
\begin{align*}
P(\gamma)\simeq & P(\gamma|T=T_{0})+\\
&
+a\int_{-\infty}^{\infty}\frac{ds}{2\pi}e^{-is\gamma}\left[e^{i\left(T_{0}+\frac{a}{2}\right)\rho s-\frac{1}{2}\frac{\rho\left(T_{0}+\frac{a}{2}\right)\left(1-\rho\left(T_{0}+\frac{a}{2}\right)\right)}{M}s^{2}}-e^{i\left(T_{0}-\frac{a}{2}\right)\rho s-\frac{1}{2}\frac{\rho\left(T_{0}-\frac{a}{2}\right)\left(1-\rho\left(T_{0}-\frac{a}{2}\right)\right)}{M}s^{2}}\right]\\
 & -\frac{a}{1+T_{0}}\int_{-\infty}^{\infty}\frac{ds}{2\pi}e^{-is\gamma}\left[e^{i\frac{\left(T_{0}+\frac{a}{2}\right)}{1+T_{0}+\frac{a}{2}}\rho s-\frac{1}{2}\frac{\rho\left(T_{0}+\frac{a}{2}\right)\left(1-\rho\left(T_{0}+\frac{a}{2}\right)\right)}{M(1+T_{0}+a/2)^{2}}s^{2}}-e^{i\frac{\left(T_{0}-\frac{a}{2}\right)}{1+T_{0}-\frac{a}{2}}\rho s-\frac{1}{2}\frac{\rho\left(T_{0}-\frac{a}{2}\right)\left(1-\rho\left(T_{0}-\frac{a}{2}\right)\right)}{M(1+T_{0}-a/2)^{2}}s^{2}}\right]\\
 & = P(\gamma|T=T_{0})+\\
 & +a\mathcal{N}\left(\rho[T_{0}+a/2],\frac{\rho[T_{0}+a/2](1-\rho[T_{0}+a/2])}{M}\right)(\gamma)\\
&
-a\mathcal{N}\left((\rho[T-a/2],\frac{\rho[T_{0}-a/2](1-\rho[T_{0}-a/2])}{M}\right)(\gamma)\\
 & -\frac{a}{1+T_{0}+a/2}\mathcal{N}\left(\rho\frac{T_{0}+a/2}{1+T_{0}+a/2},\frac{\rho[T_{0}+a/2](1-[T_{0}+a/2]\rho)}{M(1+T_{0}+a/2)^{2}}\right)(\gamma)\\
 & +\frac{a}{1+T_{0}-a/2}\mathcal{N}\left(\rho\frac{T_{0}-a/2}{1+T_{0}-a/2},\frac{\rho[T_{0}-a/2](1-[T_{0}-a/2]\rho)}{M(1+T_{0}-a/2)^{2}}\right)(\gamma) \label{eq:P_gamma_unif_approx}.
\end{align*}
%\end{widetext}

\bigskip

%%%%%%%%%  FIG.7 
\begin{figure}
\begin{centering}
\includegraphics[width=2.9in,angle=0]{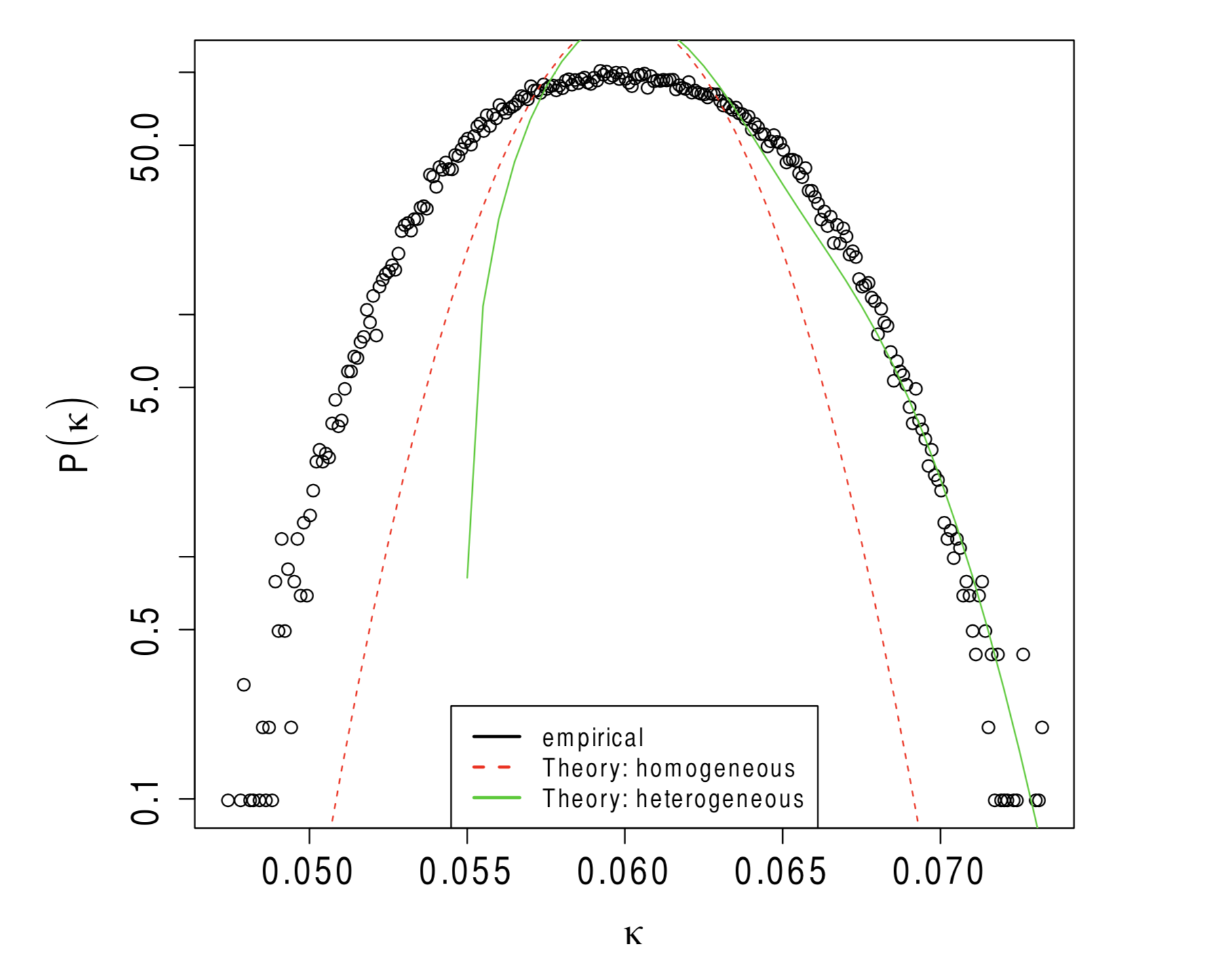} 
\caption{\label{fig:Empirical-vs-theoretical_Pkappa_heterog}Heterogeneous case: empirical (circles) vs
theoretical (red continuous line) $P(\gamma)$; $\rho=0.1$, 
$M=10^{5}$, $T\in[0.6-a/2,0.6+a/2]$ and $a=0.1$; numerical results have been obtained by Monte Carlo simulations of Eq. (\ref{eq:Pk_sum}).}
\end{centering}
\end{figure}

%%%%%%%%%  

Figure \ref{fig:Empirical-vs-theoretical_Pkappa_heterog} shows that: i) the approximation is globally too coarse, but better than the homogeneous case, and fairly good in the tails; ii) the effect of heterogeneity of talent is to make $P(\gamma)$ wider; iii) locally, the superposition (mixture) of Gaussian distributions may approximate an exponential over a given range of $\gamma$ (which must be multiplied by $M$), which may lead to a power-law part of $P(C)$. 

Alternatively, by reverting the point of view, is it possible to find the distribution of $P(T)$ that yields an approximately exponential distribution of 
$\gamma$? The answer is simple: assuming that one wishes to obtain $P(\gamma)=\lambda e^{-\lambda \gamma}$ for $\gamma\in[0,1]$, one sets
$\lambda e^{-\lambda \gamma}=\int_0^1 dT P(\gamma|T)P(T)$ and, since Eq.\ (\ref{eq:Pk_sum}) shows that $P(\gamma|T)$ is approximatively a sum of two Gaussian distributions which become very peaked and tend to a Dirac function for large $M$, one has

\begin{align}
P(\gamma)=&\lambda e^{-\lambda \gamma}\simeq \int_0^1 dT P(T) \mathcal{N}\left(\rho T,\frac{\rho T(1-\rho T)}{M}\right)(\gamma)\\&-\frac{1}{1+T}\mathcal{N}\left(\rho\frac{T}{1+T},\frac{\rho T(1-T\rho)}{M(1+T)^{2}}\right)(\gamma)\nonumber \\ 
\simeq&\int_0^1 dT P(T) \left[\delta\left (\rho T-\gamma\right )\right.\nonumber\\
&~~~~~~~~~~~~~~~~~-\left.\dfrac{1}{1+T}\delta \left (\rho \dfrac{T}{1+T}-\gamma\right )
\right ].
\end{align}

The first Dirac selects $T=\gamma/\rho$ and the second one $T=\gamma/(\rho-\gamma)$, thus
\begin{equation}
P(\gamma)\simeq \frac{1}{\rho}P(T=\gamma/\rho)-\frac{1}{|\rho-\gamma|}P[T=\gamma/(\rho-\gamma)] .
\end{equation}

If $P(T)\propto e^{-T/\rho}$, the second term becomes negligible when $\rho-\gamma\ll 1$, i.e., for the large $\gamma$ region from which the tails of $P(C)$ originate. Thus the leading contribution to $P(\gamma)$ and $P(C)$ comes from the first term of the right hand side of the above equation, which leads to roughly exponentially-distributed $\gamma$ and, mechanically, to an approximately power-law distributed $C$.
\section{Conclusions}

The main reason for the emergence of heavy-tails in the simplified Talent vs Luck model is the coupling of a multiplicative process for the capital dynamics  and a stochastic occurrence of lucky events. However, the emergence of non-trivial distributions of capital comes from the doubly stochastic nature of transformation of lucky events into a capital-increasing events: when talent is distributed homogeneously among the agent population and in the large time limit, the capital distribution is log-normal only when talent equals 1 for all the agents, i.e., when the transformation of lucky events is not stochastic. Quite remarkably, a small heterogeneity in talent makes the final distribution of capital much more complex that a simple log-normal distribution, similar to a power-law for a given range of capital, but with a truncation. This means that non-trivial heavy-tailed capital distributions do not emerge through interaction, as in e.g. \cite{Bouchad,Garlaschelli,Patriarca,Fiaschi}, but via an original two-layer stochastic process.

Extending this result to the original Talent vs Luck model is relatively simple (see appendix A), as the latter adds the occurrence of negative events, whose effect would be to reduce the effective number of lucky events occurring to each individual and to add some more noise. On the other hand, the complexity of the analytical approach increases even further in the most general case, with greater talent heterogeneity, thus making the task of finding a formal analytical relationship between the distributions of capital, talent and luck in either the TvL or the STvL models a really hard problem. 

\section*{Acknowledgements}
A.P. and A.R. acknowledge financial support by  the project "Linea di intervento 2" of the  Department of Physics and Astronomy {\it Ettore Majorana} of the University of Catania. A.P., A.E.B. and A.R. also acknowledge financial support of PRIN 2017 "Stochastic forecasting in complex systems".

\appendix

\section{Extension to the original TvL model}

In this appendix we extend the analytical approach proposed in the main part of this paper to the original TvL model. The difference is that here we also include bad luck: there is a number $N_B$ of bad luck events that follow the same dynamics as the lucky ones (independent diffusion), or equivalently, a density $\omega$ of bad luck events. Whenever a bad luck event touches an agent, this decreases the capital of the latter by a factor 2, irrespective of the talent of the said agent. Let us denote by $b_i$ the number of bad luck events that have happened to agent $i$ during the whole simulation, i.e., until time $M$: as before, we need to distinguish the number $n_i$ of lucky events and the number $k_i$ of lucky events transformed into a capital increase, each with probability $T_i$. Let us drop once again the indices $i$. The probability distribution of $b$ is $P(b)=\mathcal{N}(\mu_b,\sigma^2_b)$ with $\mu_b = \omega M$ and $\sigma^2_b =M\omega(1-\omega)$. The net luck is $D=k-b$ and is given by
\begin{equation}
P(D)=\sum_{k,b}\delta_{D,k-b}P(k)P(b). \label{eq:P_D_discrete}
\end{equation}
It is advantageous to take the large $M$ limit: $\gamma=k/M$, $\nu=n/M$, $\beta=b/M$, and $\Delta=D/M$. This leads to the continuous approximation
\begin{equation}
P(\Delta)\simeq \int_0^1 d\gamma\ d\beta\ P(\gamma)P(\beta) \delta[\Delta -(\gamma-\beta)],\label{eq:P_Delta_continuous}
\end{equation}
where $P(\gamma)$ is given by Eq.\ \eqref{eq:Pk} and $P(\beta)\simeq \mathcal{N}[\omega,\omega(1-\omega)/M](\beta)$

\subsection{The homogeneous case $T=1$}

In this case, $P(\gamma)=P(\nu)$ is a Gaussian; since $\Delta$ is the difference of $\gamma$ and $\beta$, two Gaussian variables,  $$
P(\Delta)\simeq\mathcal{N}(\rho-\omega,[\omega(1-\omega)+\rho(1-\rho)]/M)(\Delta)$$
 for $M\gg 1$. Since $C=2^{ M \Delta}$, $P(C)$ is a log-normal, as before. Interestingly, if $\rho<\omega$, i.e., if the expected number of unlucky events is greater than that of lucky events, the resulting log-normal distributions are even harder to distinguish from power law ones, adding a twist to the insights given by the TvL model.

\subsection{The homogeneous case $T<1$}

Since the lucky events are independent from the unlucky ones, Eq.(6) still holds and can be plugged into Eq.\ \eqref{eq:P_Delta_continuous}, which yields
\begin{align}
P(\Delta)  \simeq&\ \mathcal{N}\left(\rho T-\omega,\frac{\rho T(1-\rho T)+\omega(1-\omega)}{M}\right)(\Delta)\\&-\frac{1}{1+T}\mathcal{N}\left(\rho\frac{T}{1+T}-\omega,\frac{\rho T(1-T\rho)+\omega(1-\omega)(1+T)^2}{M(1+T)^{2}}\right)(\Delta). \label{eq:P_Delta_T_lt_1} 
\end{align}
Hence, the contribution of bad luck is to shift the distribution of $\Delta$ and to add some more noise, independently of $T$.

\subsection{Heterogenous talent}

The same difficulty as for the STvL model arises here: the richness of the model comes from the way heterogeneous talent spreads $P(\Delta)$, but it is out of reach of exact analytical approaches. 
Generalizing results from the STvL is the same as above: one can plug Eq.\ \eqref{eq:P_gamma_unif} into Eq.\ \eqref{eq:P_Delta_continuous} and use the approximation given by Eq.\ \eqref{eq:P_gamma_unif_approx}. Then, since $\Delta=\gamma-\beta$, Gaussian approximations allow to derive to an approximation of $P(\Delta)$. As above, the contribution of bad luck is to shift the distribution of $\Delta$ and to add some more noise, independently of $P(T)$.


\begin{thebibliography}{99}

\bibitem{Pluchino1}
Pluchino, A., Biondo, A. E., Rapisarda, A. "Talent vs Luck: the role of randomness in success and failure". Advances in Complex Systems - Vol.21, No.03n04, 1850014 (2018).

\bibitem{Hardoon}
Hardoon, D. "An economy for the 99\%", Oxfam GB, Oxfam House, John Smith Drive, Cowley, Oxford, OX4 2JY, UK (2017).

\bibitem{Pareto}
Pareto, V. "Cours d'Economique Politique", vol. 2 (1897).

\bibitem{Bouchad}
Bouchaud, J. P., \& M\'ezard, M. "Wealth condensation in a simple model of economy". Physica A: Statistical Mechanics and its Applications, 282(3-4), 536-545 (2000).

\bibitem{Garlaschelli}
Garlaschelli, D., \& Loffredo, M. I. "Effects of network topology on wealth distributions". Journal of Physics A: Mathematical and Theoretical, 41(22), 224018 (2008).

\bibitem{Fiaschi}
Fiaschi, D., \& Marsili, M. "Economic interactions and the distribution of wealth". In "Econophysics and economics of games, social choices and quantitative techniques" (pp. 61-70). Springer, Milano  (2010).

\bibitem{Patriarca}
Patriarca, M., Heinsalu, E., \& Chakraborti, A. "Basic kinetic wealth-exchange models: common features and open problems". The European Physical Journal B, 73(1), 145-153 (2010).

\bibitem{Angle}
Angle, J. (1986). The surplus theory of social stratification and the size distribution of personal wealth. Social Forces, 65(2), 293-326.

\bibitem{LuxAngle}
Lux, T. (2005). Emergent statistical wealth distributions in simple monetary exchange models: a critical review. In Econophysics of wealth distributions (pp. 51-60). Springer, Milano.

\bibitem{Wechsler}
Wechsler, D. "The Measurement and Appraisal of Adult Intelligence (fourth ed.)", Baltimore (MD): Williams and Witkins (1958).

\bibitem{Kaufman1}
Kaufman, A. S. "Assessing Adolescent and Adult Intelligence (first ed.)", Boston (MA): Allyn and Bacon (1990). 

\bibitem{Kaufman2}
Kaufman, A. S. "IQ Testing 101", New York: Springer Publishing (2009).

\bibitem{Stewart}
Stewart, J. "The Distribution of Talent", Marilyn Zurmuehlin Working Papers in Art Education 2 (1983): 21-22.

\bibitem{Erickson}
See for example: Erickson, C. "Sustainable pace is a smart long-term strategy", https://greatnotbig.com/2016/05/sustainable-pace/ (2016).

\bibitem{Milanovic}
Milanovic, B. "Global inequality of opportunity: How much of our income is determined by where we live?", Rev. Econ. Stat. 97(2) (2015) 452-460.

\bibitem{Merton}
Merton, R. K. "The Matthew effect in science", Science 159 (1968) 56-63.
	
\bibitem{Netlogo}
Wilensky, U. NetLogo. http://ccl.northwestern.edu/netlogo/. Center for Connected Learning and Computer-Based Modeling, Northwestern University, Evanston, IL (1999).

\bibitem{Clauset}
Clauset, A., Shalizi, C. R., Newman, M. E. "Power-law distributions in empirical data". SIAM review, 51(4), 661-703 (2009).

\bibitem{Voitalov}
Voitalov, I., van der Hoorn, P., van der Hofstad, R., and  Krioukov, D. (2018). Scale-free networks well done. arXiv preprint arXiv:1811.02071.

\bibitem{powerlawR}

Gillespie, C. S. "Fitting heavy tailed distributions: the poweRlaw package." arXiv preprint arXiv:1407.3492 (2014).

\bibitem{powerlawPython}
Alstott, J., Bullmore, E. \& Plenz, D. "Powerlaw: a Python package for analysis of heavy-tailed distributions". PloS one, 9(1), e85777  (2014)

\end{thebibliography}
\end{document}